\documentclass[12pt,aps,prb]{revtex4}
\usepackage{natbib,graphicx,amssymb,amsfonts}
\newcommand{\bnabla}{\nabla}

\begin{document}

\title{Electronic structure methods: Augmented Waves, Pseudopotentials
and the Projector Augmented Wave Method}

\author{Peter E. Bl\"ochl$^1$, Johannes K\"astner$^1$, and 
Clemens J. F\"orst$^{1,2}$}

\affiliation{$^1$ Clausthal University of Technology, Institute for
Theoretical Physics, Leibnizstr.10, D-38678 Clausthal-Zellerfeld,
Germany}
\affiliation{$^2$ Vienna University of Technology, Institute for
Materials Chemistry, Getreidemarkt 9/165-TC, A-1060 Vienna, Austria}
\date{\today}

\begin{abstract}

The main goal of electronic structure methods is to solve the
Schr\"odinger equation for the electrons in a molecule or solid, to
evaluate the resulting total energies, forces, response functions and
other quantities of interest.  In this paper we describe the basic ideas
behind the main electronic structure methods such as the pseudopotential
and the augmented wave methods and provide selected pointers to
contributions that are relevant for a beginner. We give particular
emphasis to the Projector Augmented Wave (PAW) method developed by one
of us, an electronic structure method for ab-initio molecular dynamics
with full wavefunctions. We feel that it allows best to show the common
conceptional basis of the most widespread electronic structure methods
in materials science.

\end{abstract}

\maketitle

\section{Introduction}

The methods described below require as input only the charge and mass
of the nuclei, the number of electrons and an initial atomic
geometry. They predict binding energies accurate within a few tenths
of an electron volt and bond-lengths in the 1-2 percent
range. Currently, systems with few hundred atoms per unit cell can be
handled. The dynamics of atoms can be studied up to tens of
pico-seconds. Quantities related to energetics, the atomic structure
and to the ground-state electronic structure can be extracted.

In order to lay a common ground and to define some of the symbols, let
us briefly touch upon the density functional theory
\citep{Kohn64,Kohn65}. It maps a description for interacting electrons, a
nearly intractable problem, onto one of non-interacting electrons in
an effective potential. Within density functional theory, the total
energy is written as

\begin{eqnarray}
E[\Psi_n(\mathbf{r}),\mathbf{R}_R]&=&\sum_n f_n 
\langle\Psi_n|\frac{-\hbar^2}{2m_e}\mathbf{\bnabla}^2|\Psi_n\rangle
\nonumber\\
&&+\frac{1}{2}\cdot\frac{e^2}{4\pi\epsilon_0}\int d^3r\int
d^3r'\;\frac{\left(n(\mathbf{r})+Z(\mathbf{r})\right)
\left(n(\mathbf{r}')+Z(\mathbf{r}')\right)}{|\mathbf{r}-\mathbf{r}'|}
+E_{xc}[n(\mathbf{r})]
\label{eq:dfttotalenergy}
\end{eqnarray}

Here, $|\Psi_n\rangle$ are one-particle electron states, $f_n$ are the
state occupations, $n(\mathbf{r})=\sum_n
f_n\Psi_n^*(\mathbf{r})\Psi_n(\mathbf{r})$ is the electron density and
$Z(\mathbf{r})=- \sum_R \mathcal{Z}_R\delta(\mathbf{r}-\mathbf{R}_R)$
is the nuclear charge density density expressed in electron charges.
$\mathcal{Z}_R$ is the atomic number of a nucleus at position
$\mathbf{R}_R$. It is implicitly assumed that the infinite
self-interaction of the nuclei is removed.  The exchange and
correlation functional contains all the difficulties of the
many-electron problem. The main conclusion of the density functional
theory is that $E_{xc}$ is a functional of the density.

We use Dirac's bra and ket notation. A wavefunction $\Psi_n$
corresponds to a ket $|\Psi_n\rangle$, the complex conjugate wave
function $\Psi_n^*$ corresponds to a bra $\langle\Psi_n|$, and a
scalar product $\int d^3r \Psi_n^*(\mathbf{r})\Psi_m(\mathbf{r})$ is
written as $\langle\Psi_n|\Psi_m\rangle$. Vectors in the 3-d
coordinate space are indicated by boldfaced symbols.  Note that we use
$\mathbf{R}$ as position vector and $R$ as atom index.

In current implementations, the exchange and correlation functional
$E_{xc}[n(\mathbf{r})]$ has the form
\begin{eqnarray*}
E_{xc}[n(\mathbf{r})]&=&\int d^3r\; 
F_{xc}(n(\mathbf{r}),|\mathbf{\bnabla}{n}(\mathbf{r})|)
\end{eqnarray*}
where $F_{xc}$ is a parameterized function of the density and its
gradients.  Such functionals are called gradient corrected.  In local
spin density functional theory, $F_{xc}$ furthermore depends on the
spin density and its derivatives. A review of the earlier developments
has been given by Parr \citep{Parr89}.

The electronic ground state is determined by minimizing the total
energy functional $E[\Psi_n]$ of Eq.~\ref{eq:dfttotalenergy} at a
fixed ionic geometry. The one-particle wavefunctions have to be
orthogonal. This constraint is implemented with the method of Lagrange
multipliers. We obtain the ground state wavefunctions from the
extremum condition for
\begin{eqnarray}
F[\Psi_n(\mathbf{r}),\Lambda_{m,n}]=E[\Psi_n]
-\sum_{n,m}[\langle\Psi_n|\Psi_m\rangle-\delta_{n,m}]\Lambda_{m,n}
\label{eq:ewithconstraint}
\end{eqnarray}
with respect to the wavefunctions and the Lagrange multipliers
$\Lambda_{m,n}$.  The extremum condition for the wavefunctions has
the form
\begin{eqnarray}
H|\Psi_n\rangle f_n =\sum_m|\Psi_m\rangle\Lambda_{m,n}
\end{eqnarray}
where
$H=-\frac{\hbar^2}{2m_e}\mathbf{\bnabla}^2+v_{\mathrm{eff}}(\mathbf{r})$
is the effective one-particle Hamilton operator.  The effective
potential depends itself on the electron density via
\begin{eqnarray*}
v_{eff}(\mathbf{r})&=&\frac{e^2}{4\pi\epsilon_0}\int
d^3r'\frac{n(\mathbf{r}')+
Z(\mathbf{r}')}{|\mathbf{r}-\mathbf{r}'|}
+\mu_{xc}(\mathbf{r})
\end{eqnarray*}
where $\mu_{xc}(\mathbf{r})=\frac{\delta E_{xc}[n(\mathbf{r})]}{\delta
n(\mathbf{r})}$ is the functional derivative of the exchange and correlation
functional.

After a unitary transformation that diagonalizes the matrix of Lagrange
multipliers $\Lambda_{m,n}$, we obtain the Kohn-Sham equations.
\begin{eqnarray}
H|\Psi_n\rangle=|\Psi_n\rangle\epsilon_n
\label{eq:hpsigleichepsi}
\end{eqnarray}
The one-particle energies $\epsilon_n$ are the eigenvalues of
$\Lambda_{n,m}\frac{f_n+f_m}{2f_nf_m}$ \citep{PAW94}.

The remaining one-electron Schr\"odinger equations, namely the
Kohn-Sham equations given above, still pose substantial numerical
difficulties: (1) in the atomic region near the nucleus, the kinetic
energy of the electrons is large, resulting in rapid oscillations of
the wavefunction that require fine grids for an accurate numerical
representation.  On the other hand, the large kinetic energy makes the
Schr\"odinger equation stiff, so that a change of the chemical
environment has little effect on the shape of the wavefunction.
Therefore, the wavefunction in the atomic region can be represented
well already by a small basis set.  (2) In the bonding region between
the atoms the situation is opposite.  The kinetic energy is small and
the wavefunction is smooth.  However, the wavefunction is flexible
and responds strongly to the environment. This requires large and
nearly complete basis sets.

Combining these different requirements is non-trivial and various
strategies have been developed.
\begin{itemize}

\item The atomic point of view has been most appealing to quantum
chemists.  Basis functions that resemble atomic orbitals are
chosen. They exploit that the wavefunction in the atomic region can
be described by a few basis functions, while the chemical bond is
described by the overlapping tails of these atomic orbitals.  Most
techniques in this class are a compromise of, on the one hand, a
well adapted basis set, where the basis functions are difficult to
handle, and on the other hand numerically convenient basis functions
such as Gaussians, where the inadequacies are compensated by larger
basis sets.

\item Pseudopotentials regard an atom as a perturbation of the free
electron gas. The most natural basis functions are plane-waves. Plane
waves basis sets are in principle complete and suitable for sufficiently
smooth wavefunctions. The disadvantage of the comparably large basis
sets required is offset by their extreme numerical simplicity. Finite
plane-wave expansions are, however, absolutely inadequate to describe
the strong oscillations of the wavefunctions near the nucleus.  In the
pseudopotential approach the Pauli repulsion of the core electrons is
therefore described by an effective potential that expels the valence
electrons from the core region.  The resulting wavefunctions are smooth
and can be represented well by plane-waves.  The price to pay is that
all information on the charge density and wavefunctions near the
nucleus is lost.

\item Augmented wave methods compose their basis functions from
atom-like wavefunctions in the atomic regions and a set of
functions, called envelope functions, appropriate for the bonding in
between.  Space is divided accordingly into atom-centered spheres,
defining the atomic regions, and an interstitial region in between.
The partial solutions of the different regions, are matched at the
interface between atomic and interstitial regions.

\end{itemize}
The projector augmented wave method is an extension of
augmented wave methods and the pseudopotential approach, which
combines their traditions into a unified electronic structure method.

After describing the underlying ideas of the various approaches let us
briefly review the history of augmented wave methods and the
pseudopotential approach. We do not discuss the atomic-orbital based
methods, because our focus is the PAW method and its ancestors.

\section{Augmented Wave Methods}

The augmented wave methods have been introduced in 1937 by Slater
\citep{Slater37} and were later modified by Korringa
\citep{Korringa47}, Kohn and Rostokker \citep{Kohn54}. They approached
the electronic structure as a scattered-electron problem. Consider an
electron beam, represented by a plane-wave, traveling through a solid.
It undergoes multiple scattering at the atoms.  If for some energy,
the outgoing scattered waves interfere destructively, a bound state
has been determined.  This approach can be translated into a basis set
method with energy and potential dependent basis functions.  In order
to make the scattered wave problem tractable, a model potential had to
be chosen: The so-called muffin-tin potential approximates the
true potential by a constant in the interstitial region and by a
spherically symmetric potential in the atomic region.

Augmented wave methods reached adulthood in the 1970s: O.K.~Andersen
\citep{Andersen75} showed that the energy dependent basis set of
Slater's APW method can be mapped onto one with energy independent
basis functions, by linearizing the partial waves for the atomic
regions in energy.  In the original APW approach, one had to
determine the zeros of the determinant of an energy dependent matrix,
a nearly intractable numerical problem for complex systems.  With the
new energy independent basis functions, however, the problem is
reduced to the much simpler generalized eigenvalue problem, which can
be solved using efficient numerical techniques. Furthermore, the
introduction of well defined basis sets paved the way for
full-potential calculations \citep{Krakauer79}. In that case the
muffin-tin approximation is used solely to define the basis set
$|\chi_i\rangle$, while the matrix elements
$\langle\chi_i|H|\chi_j\rangle$ of the Hamiltonian are evaluated with
the full potential.

In the augmented wave methods one constructs the basis set for the
atomic region by solving the radial Schr\"odinger equation for the
spheridized effective potential
\begin{eqnarray*}
\left[\frac{-\hbar^2}{2m_e}\mathbf{\bnabla}^2
+v_{eff}(\mathbf{r})-\epsilon\right]
\phi_{\ell,m}(\epsilon,\mathbf{r})&=&0
\end{eqnarray*}
as function of energy. Note that a partial wave
$\phi_{\ell,m}(\epsilon,\mathbf{r})$ is an angular momentum eigenstate
and can be expressed as a product of a radial function and a spherical
harmonic. The energy dependent partial wave is expanded in a Taylor
expansion about some reference energy $\epsilon_{\nu,\ell}$
\begin{eqnarray*}
\phi_{\ell,m}(\epsilon,\mathbf{r})&=&\phi_{\nu,\ell,m}(\mathbf{r})
+(\epsilon-\epsilon_{\nu,\ell})\dot\phi_{\nu,\ell,m}(\mathbf{r})
+O((\epsilon-\epsilon_{\nu,\ell})^2)
\end{eqnarray*}
where
$\phi_{\nu,\ell,m}(\mathbf{r})=\phi_{\ell,m}(\epsilon_{\nu,\ell},\mathbf{r})$.
The energy derivative of the partial wave 
$\dot{\phi}_\nu(\mathbf{r})
=\left.\frac{\partial\phi(\epsilon,\mathbf{r})}
{\partial\epsilon}\right|_{\epsilon_{\nu,\ell}}$ solves the equation
\begin{eqnarray*}
\left[\frac{-\hbar^2}{2m_e}\mathbf{\bnabla}^2
+v_{eff}(\mathbf{r})-\epsilon_{\nu,\ell}\right]
\dot\phi_{\nu,\ell,m}(\mathbf{r})
&=&\phi_{\nu,\ell,m}(\mathbf{r})
\end{eqnarray*}

Next one starts from a regular basis set, such as plane-waves,
Gaussians or Hankel functions. These basis functions are called
envelope functions $|\tilde\chi_i\rangle$.  Within the atomic region
they are replaced by the partial waves and their energy derivatives,
such that the resulting wavefunction is continuous and
differentiable.
\begin{eqnarray}
\chi_i(\mathbf{r})
=\tilde\chi_i(\mathbf{r})-\sum_R \theta_R(\mathbf{r})
\tilde\chi_i(\mathbf{r})
+\sum_{R,\ell,m}\theta_R(\mathbf{r})
\left[\phi_{\nu,R,\ell,m}(\mathbf{r})a_{R,\ell,m,i} 
+\dot\phi_{\nu,R,\ell,m}(\mathbf{r})b_{R,\ell,m,i} \right]
\label{eq:augmentedwave}
\end{eqnarray}
$\theta_R(\mathbf{r})$ is a step function that is unity within the
augmentation sphere centered at $\mathbf{R}_R$ and zero elsewhere.
The augmentation sphere is atom-centered and has a radius about equal
to the covalent radius. This radius is called the muffin-tin radius,
if the spheres of neighboring atoms touch.  These basis functions
describe only the valence states; the core states are localized within
the augmentation sphere and are obtained directly by radial
integration of the Schr\"odinger equation within the augmentation
sphere.

The coefficients $a_{R,\ell,m,i}$ and $b_{R,\ell,m,i}$ are obtained
for each $|\tilde\chi_i\rangle$ as follows: The envelope function is
decomposed around each atomic site into spherical harmonics multiplied
by radial functions. 
\begin{eqnarray}
\tilde\chi_i(\mathbf{r})
=\sum_{\ell,m}u_{R,\ell,m,i}(|\mathbf{r}-\mathbf{R}_R|)
Y_{\ell,m}(\mathbf{r}-\mathbf{R}_R)
\end{eqnarray}
Analytical expansions for plane-waves, Hankel functions or Gaussians exist.
The radial parts of the partial waves $\phi_{\nu,R,\ell,m}$ and
$\dot\phi_{\nu,R,\ell,m}$ are matched with value and derivative to
$u_{R,\ell,m,i}(|\mathbf{r}|)$, which yields the expansion
coefficients $a_{R,\ell,m,i}$ and $b_{R,\ell,m,i}$.

If the envelope functions are plane-waves, the resulting method is
called the linear augmented plane-wave (LAPW) method. If the envelope
functions are Hankel functions, the method is called linear muffin-tin
orbital (LMTO) method .

A good review of the LAPW method \citep{Andersen75} has been given by
Singh \citep{Singhbook}. Let us now briefly mention the major
developments of the LAPW method: Soler \citep{Soler89} introduced the
idea of additive augmentation: While augmented plane-waves are
discontinuous at the surface of the augmentation sphere if the
expansion in spherical harmonics in Eq.~\ref{eq:augmentedwave} is
truncated, Soler replaced the second term in
Eq.~\ref{eq:augmentedwave} by an expansion of the plane-wave with the
same angular momentum truncation as in the third term. This
dramatically improved the convergence of the angular momentum
expansion.  Singh \citep{Singh91} introduced so-called local orbitals,
which are non-zero only within a muffin-tin sphere, where they are
superpositions of $\phi$ and $\dot\phi$ functions from different
expansion energies. Local orbitals substantially increase the energy
transferability. Sj\"ostedt \citep{Sjoestedt00} relaxed the condition
that the basis functions are differentiable at the sphere radius. In
addition she introduced local orbitals, which are confined inside the
sphere, and that also have a kink at the sphere boundary. Due to the
large energy-cost of kinks, they will cancel, once the total energy is
minimized. The increased variational degree of freedom in the basis
leads to a dramatically improved plane-wave convergence
\citep{Madsen01}.  

The second variant of the linear methods is the LMTO method
\citep{Andersen75}.  A good introduction into the LMTO method is the
book by Skriver \citep{Skriverbook}. The LMTO method uses Hankel
functions as envelope functions. The atomic spheres approximation
(ASA) provides a particularly simple and efficient approach to the
electronic structure of very large systems.  In the ASA the
augmentation spheres are blown up so that their volume are equal to
the total volume and the first two terms in Eq.~\ref{eq:augmentedwave}
are ignored. The main deficiency of the LMTO-ASA method is the
limitation to structures that can be converted into a closed packed
arrangement of atomic and empty spheres. Furthermore energy
differences due to structural distortions are often qualitatively
incorrect. Full potential versions of the LMTO method, that avoid
these deficiencies of the ASA have been developed.  The construction
of tight binding orbitals as superposition of muffin-tin orbitals
\citep{Jepsen84} showed the underlying principles of the empirical
tight-binding method and prepared the ground for electronic structure
methods that scale linearly instead of with the third power of the
number of atoms.  The third generation LMTO \citep{Andersen03} allows
to construct true minimal basis sets, which require only one orbital
per electron-pair for insulators. In addition they can be made
arbitrarily accurate in the valence band region, so that a matrix
diagonalization becomes unnecessary.  The first steps towards a
full-potential implementation, that promises a good accuracy, while
maintaining the simplicity of the of the LMTO-ASA method are currently
under way.  Through the minimal basis-set construction the LMTO method
offers unrivaled tools for the analysis of the electronic structure
and has been extensively used in hybrid methods combining density
functional theory with model Hamiltonians for materials with strong
electron correlations \citep{Held02}

\section{Pseudopotentials}

Pseudopotentials have been introduced to (1) avoid describing the core
electrons explicitely and (2) to avoid the rapid oscillations of the
wavefunction near the nucleus, which normally require either
complicated or large basis sets.

The pseudopotential approach traces back to 1940 when C. Herring
invented the orthogonalized plane-wave method \citep{Herring40}. Later,
Phillips \citep{Phillips58} and Antoncik \citep{Antoncik} replaced the
orthogonality condition by an effective potential, which mimics the
Pauli-repulsion by the core electrons and thus compensates the
electrostatic attraction by the nucleus. In practice, the potential
was modified, for example, by cutting off the singular potential of
the nucleus at a certain value. This was done with a few parameters
that have been adjusted to reproduce the measured electronic band
structure of the corresponding solid.

Hamann, Schl\"uter and Chiang \citep{Hamann79} showed in 1979 how
pseudopotentials can be constructed in such a way, that their
scattering properties are identical to that of an atom to first order
in energy. These first-principles pseudopotentials relieved the
calculations from the restrictions of empirical parameters. Highly
accurate calculations have become possible especially for
semiconductors and simple metals. An alternative approach towards
first-principles pseudopotentials \citep{Zunger78} preceeded the one
mentioned above.

\subsection{The idea behind Pseudopotential construction}

In order to construct a first-principles pseudopotential, one starts out
with an all-electron density-functional calculation for a spherical
atom. Such calculations can be performed efficiently on radial grids.
They yield the atomic potential and wavefunctions
$\phi_{\ell,m}(\mathbf{r})$.  Due to the spherical symmetry, the radial
parts of the wavefunctions for different magnetic quantum numbers $m$
are identical.

\begin{figure}
\centering
\includegraphics[width=10cm,clip=true]{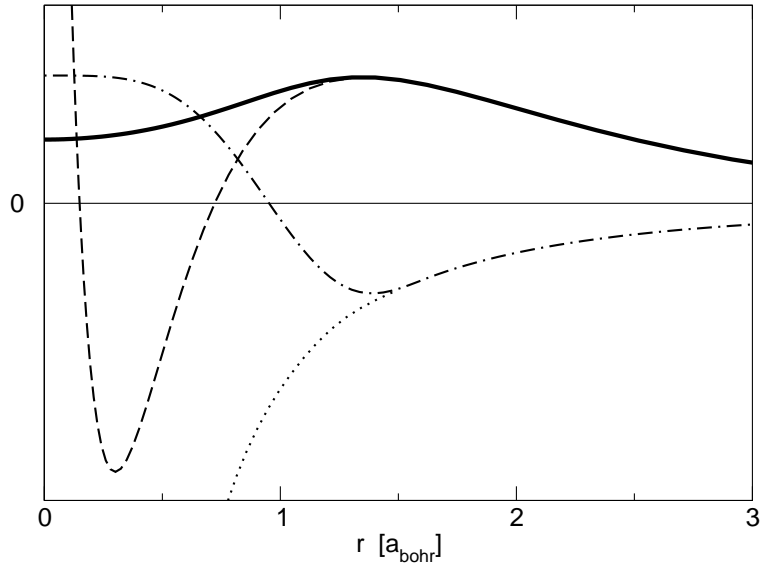}
\caption{Illustration of the pseudopotential concept at the example of
the 3$s$ wavefunction of Si. The solid line shows the radial part of
the pseudo wavefunction $\tilde\phi_{\ell,m}$. The dashed line corresponds to the
all-electron wavefunction $\phi_{\ell,m}$ which exhibits strong oscillations at small
radii. The angular momentum dependent pseudopotential $u_\ell$ (dash-dotted
line) deviates from the all-electron one $v_{eff}$ (dotted line) inside the
augmentation region. The data are generated by the fhi98PP
code \citep{Fuchs99}.}
\label{fig:pp}
\end{figure}

For the valence wavefunctions one constructs pseudo wavefunctions
$|\tilde\phi_{\ell,m}\rangle$: There are numerous ways
\citep{Kerker80,Bachelet82,Troullier93,Heine93} to construct the
pseudo wavefunctions. They must be identical to the true wave
functions outside the augmentation region, which is called core-region
in the context of the pseudopotential approach. Inside the
augmentation region the pseudo wavefunction should be node-less and
have the same norm as the true wavefunctions, that is
$\langle\tilde\phi_{\ell,m}|\tilde\phi_{\ell,m}\rangle
=\langle\phi_{\ell,m}|\phi_{\ell,m}\rangle$ 
(compare Figure~\ref{fig:pp}).

From the pseudo wavefunction, a potential $u_{\ell}(\mathbf{r})$ can be
reconstructed by inverting the respective Schr\"odinger equation.
\begin{eqnarray*}
\left[-\frac{\hbar^2}{2m_e}\mathbf{\bnabla}^2
+u_{\ell}(\mathbf{r})-\epsilon_{\ell,m}\right]\tilde\phi_{\ell,m}(\mathbf{r})=0
&\Rightarrow&
u_{\ell}(\mathbf{r})=\epsilon+\frac{1}{\tilde\phi_{\ell,m}(\mathbf{r})}
\cdot\frac{\hbar^2}{2m_e}\mathbf{\bnabla}^2\tilde\phi_{\ell,m}(\mathbf{r})
\end{eqnarray*}
This potential $u_{\ell}(\mathbf{r})$ (compare Figure~\ref{fig:pp}), which is also spherically
symmetric, differs from one main angular momentum $\ell$ to the other.

Next we define an effective pseudo Hamiltonian 
\begin{eqnarray*}
\tilde{H}_\ell=-\frac{\hbar^2}{2m_e}\mathbf{\bnabla}^2+v^{ps}_{\ell}(\mathbf{r})+
\frac{e^2}{4\pi\epsilon_0} \int
d^3r'\frac{\tilde{n}(\mathbf{r}')+\tilde{Z}(\mathbf{r}')}
{|\mathbf{r}-\mathbf{r'}|} +\mu_{xc}([n(\mathbf{r})],\mathbf{r})
\end{eqnarray*}
and determine the pseudopotentials $v^{ps}_\ell$ such that the pseudo
Hamiltonian produces the pseudo wavefunctions, that is
\begin{eqnarray}
v^{ps}_{\ell}(\mathbf{r})&=&u_{\ell}(\mathbf{r})
-\frac{e^2}{4\pi\epsilon_0}\int d^3r'\; 
\frac{\tilde{n}(\mathbf{r}')+\tilde{Z}(\mathbf{r}')}{|\mathbf{r}-\mathbf{r'}|}
-\mu_{xc}([\tilde{n}(\mathbf{r})],\mathbf{r})
\label{eq:unscreening}
\end{eqnarray}
This process is called ``unscreening''. 

$\tilde{Z}(\mathbf{r})$ mimics the charge density of the nucleus and
the core electrons. It is usually an atom-centered, spherical Gaussian
that is normalized to the charge of nucleus and core of that atom.  In
the pseudopotential approach, $\tilde{Z}_R(\mathbf{r})$ it does not
change with the potential.  The pseudo density
$\tilde{n}(\mathbf{r})=\sum_n
f_n\tilde\Psi^*_n(\mathbf{r})\tilde\Psi_n(\mathbf{r})$ is constructed
from the pseudo wavefunctions.

In this way we obtain a different potential for each angular momentum
channel.  In order to apply these potentials to a given wavefunction,
the wavefunction must first be decomposed into angular momenta.  Then
each component is applied to the pseudopotential $v^{ps}_\ell$ for
the corresponding angular momentum.

The pseudopotential defined in this way can be expressed in a semi-local form 
\begin{eqnarray}
v^{ps}(\mathbf{r},\mathbf{r}')
&=&\bar{v}(\mathbf{r})\delta(\mathbf{r}-\mathbf{r}')+
\sum_{\ell,m}
\left[Y_{\ell,m}(\mathbf{r})
\left[v^{ps}_{\ell}(\mathbf{r})-\bar{v}(\mathbf{r})\right] 
\frac{ \delta(|\mathbf{r}|-|\mathbf{r}'|)}{|\mathbf{r}|^2}
Y^*_{\ell,m}(\mathbf{r}')\right]
\label{eq:semilocal}
\end{eqnarray}
The local potential $\bar{v}(\mathbf{r})$ only acts on those angular
momentum components, not included in the expansion of the
pseudopotential construction.  Typically it is chosen to cancel the
most expensive nonlocal terms, the one corresponding to the highest
physically relevant angular momentum.

The pseudopotential is non-local as it depends on two position arguments,
$\mathbf{r}$ and $\mathbf{r}'$. The expectation values are evaluated as a
double integral
\begin{eqnarray*}
\langle\tilde\Psi|v_{ps}|\tilde\Psi\rangle
&=&\int d^3r\int d^3r'\;\tilde\Psi^*(\mathbf{r})v^{ps}(\mathbf{r},\mathbf{r}')
\tilde\Psi(\mathbf{r}')
\end{eqnarray*}

The semi-local form of the pseudopotential given in Eq.~\ref{eq:semilocal}
is computationally expensive.  Therefore, in practice one uses a
separable form of the pseudopotential
\citep{Kleinman82,Bloechl90,Vanderbilt90}.
\begin{eqnarray}
v^{ps}&\approx&\sum_{i,j} v^{ps}|\tilde\phi_i\rangle 
\left[\langle\tilde\phi_j|v^{ps}|\tilde\phi_i\rangle\right]^{-1}_{i,j} 
\langle\tilde\phi_j|v^{ps}
\label{eq:separable}
\end{eqnarray}
Thus the projection onto spherical harmonics used in the semi-local
form of Eq.~\ref{eq:semilocal} is replaced by a projection onto
angular momentum dependent functions $|v^{ps}\tilde\phi_i\rangle$.

The indices $i$ and $j$ are composite indices containing the
atomic-site index $R$, the angular momentum quantum numbers $\ell,m$
and an additional index $\alpha$. The index $\alpha$ distinguishes
partial waves with otherwise identical indices $R,\ell,m$, as more
than one partial wave per site and angular momentum is allowed.  The
partial waves may be constructed as eigenstates to the pseudopotential
$v^{ps}_\ell$ for a set of energies.

One can show that the identity of Eq.~\ref{eq:separable} holds by
applying a wavefunction $|\tilde\Psi\rangle=\sum_i|\tilde\phi_i\rangle
c_i$ to both sides.  If the set of pseudo partial waves
$|\tilde\phi_i\rangle$ in Eq.~\ref{eq:separable} is complete, the
identity is exact.  The advantage of the separable form is that
$\langle\tilde\phi v^{ps}|$ is treated as one function, so that
expectation values are reduced to combinations of simple scalar products
$\langle\tilde\phi_i v^{ps}|\tilde\Psi\rangle$.

\subsection{The Pseudopotential total energy}

The total energy of the pseudopotential method can be written in the form
\begin{eqnarray}
E&=&\sum_n f_n
\langle\tilde\Psi_n|-\frac{\hbar^2}{2m_e}\mathbf{\bnabla}^2|\tilde\Psi_n\rangle
+E_{self}+\sum_n f_n
\langle\tilde\Psi_n|v_{ps}|\tilde\Psi_n\rangle
\nonumber\\
&&+\frac{1}{2}\cdot\frac{e^2}{4\pi\epsilon_0} \int d^3r\int d^3r' 
\frac{\left[\tilde{n}(\mathbf{r})+\tilde{Z}(\mathbf{r})\right]
\left[\tilde{n}(\mathbf{r}')+\tilde{Z}(\mathbf{r}')\right]}
{|\mathbf{r}-\mathbf{r}'|}
+E_{xc}[\tilde{n}(\mathbf{r})]
\label{eq:totalenergypseudopotential}
\end{eqnarray}
The constant $E_{self}$ is adjusted such that the total energy of the
atom is the same for an all-electron calculation and the
pseudopotential calculation.

For the atom, from which it has been constructed, this construction
guarantees that the pseudopotential method produces the correct
one-particle energies for the valence states and that the wave
functions have the desired shape. 

While pseudopotentials have proven to be accurate for a large variety
of systems, there is no strict guarantee that they produce the same
results as an all-electron calculation, if they are used in a molecule
or solid. The error sources can be divided into two classes:
\begin{itemize}
\item Energy transferability problems: Even for the potential of the
reference atom, the scattering properties are accurate only in given
energy window.
\item Charge transferability problems: In a molecule or crystal, the
potential differs from that of the isolated atom.  The
pseudopotential, however, is strictly valid only for the isolated
atom.
\end{itemize}

The plane-wave basis set for the pseudo wavefunctions is defined by
the shortest wave length $\lambda=2\pi/|G|$ via the so-called plane
wave cutoff $E_{PW}=\frac{\hbar^2G_{max}^2}{2m_e}$. It is often
specified in Rydberg (1~Ry=$\frac{1}{2}$~H$\approx$13.6~eV).  The
plane-wave cutoff is the highest kinetic energy of all
basis functions. The basis-set convergence can systematically be
controlled by increasing the plane-wave cutoff.

The charge transferability is substantially improved by including a
nonlinear core correction \citep{Louie82} into the exchange-correlation
term of Eq.~\ref{eq:totalenergypseudopotential}.  Hamann
\citep{Hamann89} showed, how to construct pseudopotentials also from
unbound wavefunctions. Vanderbilt \citep{Vanderbilt90,Laasonen93}
generalized the pseudopotential method to non-normconserving
pseudopotentials, so-called ultra-soft pseudopotentials, which
dramatically improves the basis-set convergence.  The formulation of
ultra-soft pseudopotentials has already many similarities with the
projector augmented wave method. Truncated separable pseudopotentials
suffer sometimes from so-called ghost states. These are unphysical
core-like states, which render the pseudopotential useless. These
problems have been discussed by Gonze \citep{Gonze91}. Quantities such
as hyperfine parameters that depend on the full wavefunctions near
the nucleus, can be extracted approximately \citep{Hyperfine}. A
good review about pseudopotential methodology has been written by
Payne \citep{Payne92} and Singh \citep{Singhbook}.

In 1985 R. Car and M. Parrinello published the ab-initio molecular
dynamics method \citep{Car85}. Simulations of the atomic motion have
become possible on the basis of state-of-the-art electronic structure
methods. Besides making dynamical phenomena and finite temperature
effects accessible to electronic structure calculations, the ab-initio
molecular dynamics method also introduced a radically new way of
thinking into electronic structure methods.  Diagonalization of a
Hamilton matrix has been replaced by classical equations of motion for
the wavefunction coefficients.  If one applies friction, the system
is quenched to the ground state.  Without friction truly dynamical
simulations of the atomic structure are performed. Using thermostats
\citep{Nose84,Hoover85,Bloechl92,Bloechl02} simulations at constant
temperature can be performed. The Car-Parrinello method treats
electronic wavefunctions and atomic positions on an equal footing.

%
\section{Projector augmented wave method}

The Car-Parrinello method had been implemented first for the
pseudopotential approach. There seemed to be unsurmountable barriers
against combining the new technique with augmented wave methods. The
main problem was related to the potential dependent basis set used in
augmented wave methods: the Car-Parrinello method requires a well
defined and unique total energy functional of atomic positions and
basis set coefficients.  Furthermore the analytic evaluation of the
first partial derivatives of the total energy with respect to wave
functions, $H|\Psi_n\rangle$, and atomic position, the forces, must be
possible.  Therefore, it was one of the main goals of the PAW method
to introduce energy and potential independent basis sets that were as
accurate as the previously used augmented basis sets. Other
requirements have been: (1) The method should at least match the
efficiency of the pseudopotential approach for Car-Parrinello
simulations.  (2) It should become an exact theory when converged and
(3) its convergence should be easily controlled. We believe that these
criteria have been met, which explains why the PAW method becomes
increasingly wide spread today.

\subsection{Transformation theory}
At the root of the PAW method lies a transformation, that maps the
true wavefunctions with their complete nodal structure onto auxiliary
wavefunctions, that are numerically convenient.  We aim for smooth
auxiliary wavefunctions, which have a rapidly convergent plane-wave
expansion. With such a transformation we can expand the auxiliary wave
functions into a convenient basis set such as plane-waves, and
evaluate all physical properties after reconstructing the related
physical (true) wavefunctions.

Let us denote the physical one-particle wavefunctions as
$|\Psi_n\rangle$ and the auxiliary wavefunctions as
$|\tilde\Psi_n\rangle$. Note that the tilde refers to the
representation of smooth auxiliary wavefunctions and $n$ is the label
for a one-particle state and contains a band index, a $k$-point and a
spin index. The transformation from the auxiliary to the physical wave
functions is denoted by ${\cal T}$.
\begin{eqnarray}
|\Psi_n\rangle={\cal T}|\tilde{\Psi}_n\rangle
\end{eqnarray}

Now we express the constrained density functional $F$ of
Eq.~\ref{eq:ewithconstraint} in terms of our auxiliary wavefunctions
\begin{eqnarray}
F[{\cal
T}\tilde\Psi_n,\Lambda_{m,n}]
= E[{\cal T}\tilde\Psi_n] -\sum_{n,m}[\langle\tilde\Psi_n|{\cal
T}^\dagger{\cal T}|\tilde\Psi_m\rangle -\delta_{n,m}]\Lambda_{m,n}
\end{eqnarray}
The variational principle with respect to the auxiliary wavefunctions
yields
\begin{eqnarray}
{\cal T}^\dagger H{\cal T}|\tilde\Psi_n\rangle
={\cal T}^\dagger{\cal T}|\tilde\Psi_n\rangle\epsilon_n.
\end{eqnarray}
Again we obtain a Schr\"odinger-like equation (see derivation of
Eq.~\ref{eq:hpsigleichepsi}), but now the Hamilton
operator has a different form, $\tilde{H}={\cal T}^\dagger H{\cal T}$,
an overlap operator $\tilde{O}={\cal T}^\dagger{\cal T}$ occurs, and
the resulting auxiliary wavefunctions are smooth.

When we evaluate physical quantities we need to evaluate expectation values of
an operator $A$, which can be expressed in terms of either the true or the
auxiliary wavefunctions.
\begin{eqnarray}
\langle A\rangle&=&\sum_nf_n\langle\Psi_n|A|\Psi_n\rangle
=\sum_nf_n\langle\tilde\Psi_n|{\cal T}^\dagger A{\cal T}|\tilde\Psi_n\rangle
\end{eqnarray}
In the representation of auxiliary wavefunctions we need to use transformed
operators $\tilde{A}={\cal T}^\dagger A{\cal T}$.  As it is, this equation
only holds for the valence electrons.  The core electrons are treated
differently as will be shown below.

The transformation takes us conceptionally from the world of pseudopotentials
to that of augmented wave methods, which deal with the full wavefunctions. We
will see that our auxiliary wavefunctions, which are simply the plane-wave
parts of the full wavefunctions, translate into the wavefunctions of the
pseudopotential approach.  In the PAW method the auxiliary wavefunctions are
used to construct the true wavefunctions and the total energy functional is
evaluated from the latter.
Thus it provides the missing link between augmented wave methods and the
pseudopotential method, which can be derived as a well-defined approximation
of the PAW method.

In the original paper \citep{PAW94}, the auxiliary wavefunctions have
been termed pseudo wavefunctions and the true wavefunctions have
been termed all-electron wavefunctions, in order to make the
connection more evident.  We avoid this notation here, because it
resulted in confusion in cases, where the correspondence is not
clear-cut.
\subsection{Transformation operator}
Sofar, we have described how we can determine the auxiliary wave
functions of the ground state and how to obtain physical information
from them.  What is missing, is a definition of the transformation
operator ${\cal T}$.

The operator ${\cal T}$ has to modify the smooth auxiliary wave
function in each atomic region, so that the resulting wavefunction
has the correct nodal structure.  Therefore, it makes sense to write
the transformation as identity plus a sum of atomic contributions
${\cal S}_R$
\begin{eqnarray}
{\cal T}=1+\sum_R{\cal S}_R.
\end{eqnarray}
For every atom, ${\cal S}_R$ adds the difference between the
true and the auxiliary wavefunction. 

The local terms ${\cal S}_R$ are defined in terms of
solutions $|\phi_{i}\rangle$ of the Schr\"odinger equation for the
isolated atoms.  This set of partial waves $|\phi_{i}\rangle$ will
serve as a basis set so that, near the nucleus, all relevant valence
wavefunctions can be expressed as superposition of the partial waves
with yet unknown coefficients.
\begin{eqnarray}
\Psi(\mathbf{r})=\sum_{i\in R}\phi_{i}(\mathbf{r}) c_i\quad {\rm
for}\quad |\mathbf{r}-\mathbf{R}_R|<r_{c,R}
\end{eqnarray}
With $i\in R$ we indicate those partial waves that belong to site
$R$.

Since the core wavefunctions do not spread out into the neighboring
atoms, we will treat them differently. Currently we use the
frozen-core approximation, which imports the density and the energy of
the core electrons from the corresponding isolated atoms. The
transformation ${\cal T}$ shall produce only wavefunctions orthogonal
to the core electrons, while the core electrons are treated
separately.  Therefore, the set of atomic partial waves
$|\phi_i\rangle$ includes only valence states that are orthogonal to
the core wavefunctions of the atom.

For each of the partial waves we choose an auxiliary partial wave
$|\tilde\phi_i\rangle$. The identity
\begin{eqnarray}
|\phi_i\rangle&=&(1+{\cal S}_R)|\tilde\phi_i\rangle
\quad\mathrm{for}\quad i\in R 
\nonumber\\
{\cal S}_R|\tilde\phi_i\rangle&=&|\phi_i\rangle-|\tilde\phi_i\rangle
\label{eq:sr1}
\end{eqnarray}
defines the local contribution ${\cal S}_R$ to the transformation
operator. Since $1+{\cal S}_R$ shall change the wavefunction only
locally, we require that the partial waves $|\phi_i\rangle$ and their
auxiliary counter parts $|\tilde\phi_i\rangle $ are pairwise identical
beyond a certain radius $r_{c,R}$.
\begin{eqnarray}
\phi_i(\mathbf{r})
=\tilde\phi_i(\mathbf{r})
\quad\mathrm{for}\quad i\in R\quad\mathrm{and}\quad|\mathbf{r}-\mathbf{R}_R|
>r_{c,R}
\label{eq:equaloutside}
\end{eqnarray}

Note that the partial waves are not necessarily bound states and are
therefore not normalizable, unless we truncate them beyond a certain
radius $r_{c,R}$. The PAW method is formulated such that the final
results do not depend on the location where the partial waves are
truncated, as long as this is not done too close to the nucleus and
identical for auxiliary and all-electron partial waves.

\begin{figure}[ht]
\centering
\includegraphics[width=2.5cm,angle=-90,clip=true]{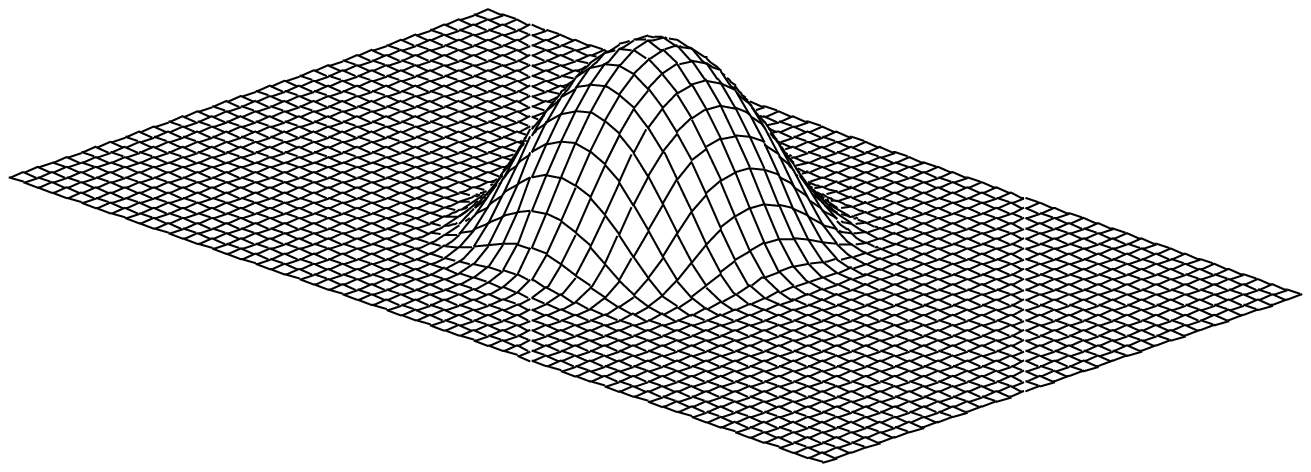}
\includegraphics[width=2.5cm,angle=-90,clip=true]{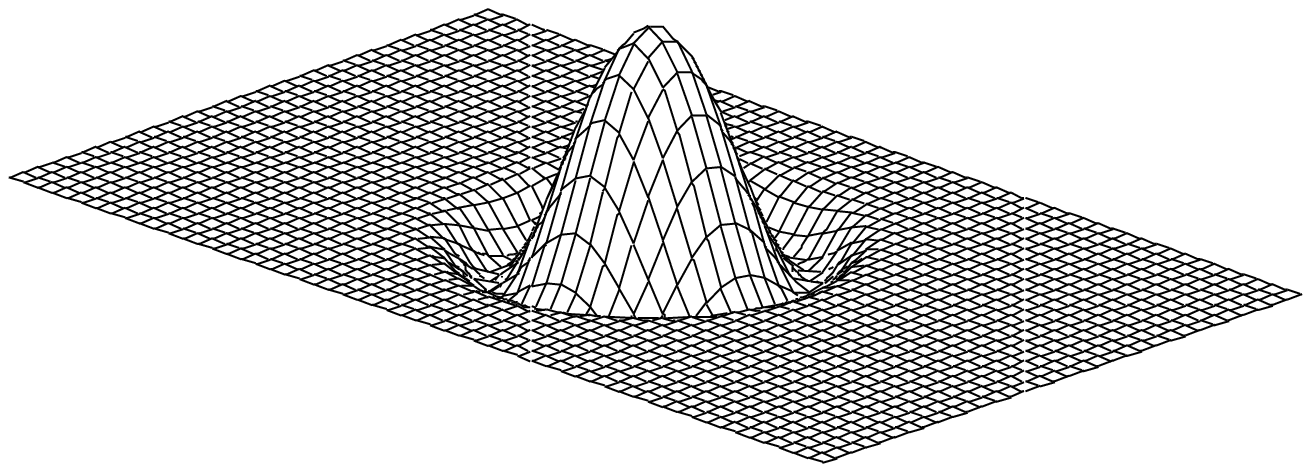}\\
\includegraphics[width=2.5cm,angle=-90,clip=true]{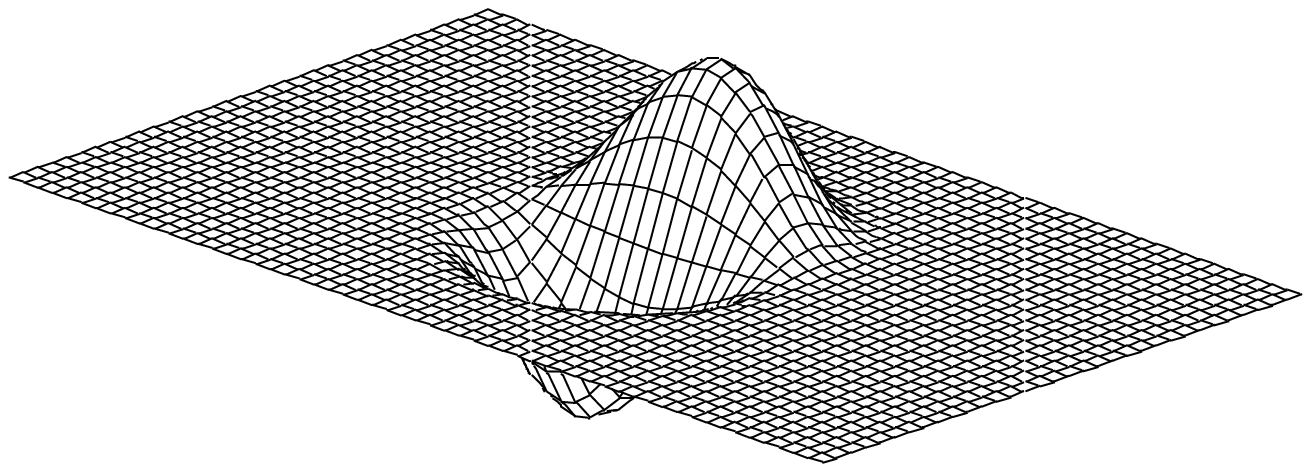}
\includegraphics[width=2.5cm,angle=-90,clip=true]{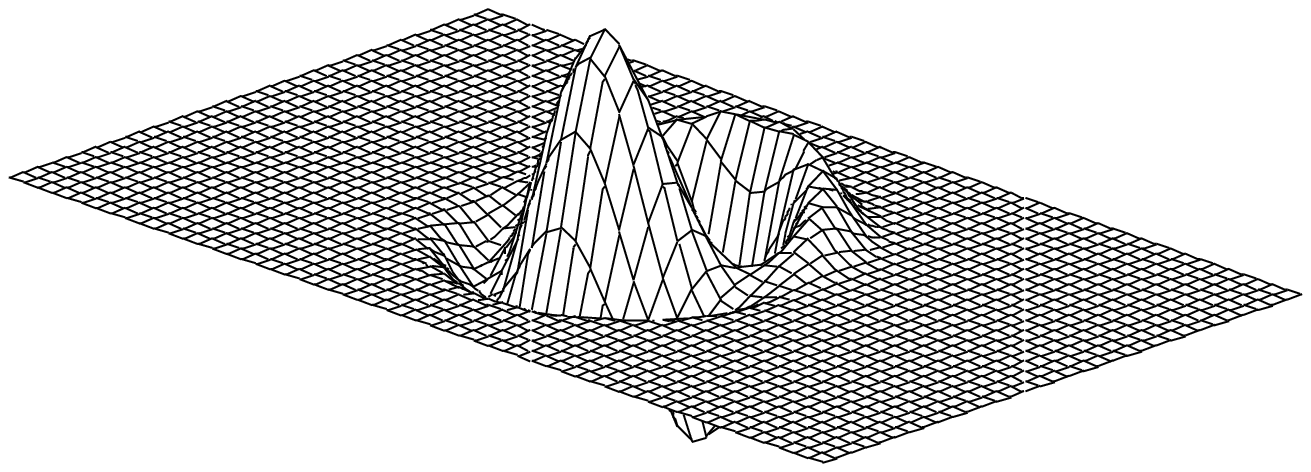}\\
\includegraphics[width=2.5cm,angle=-90,clip=true]{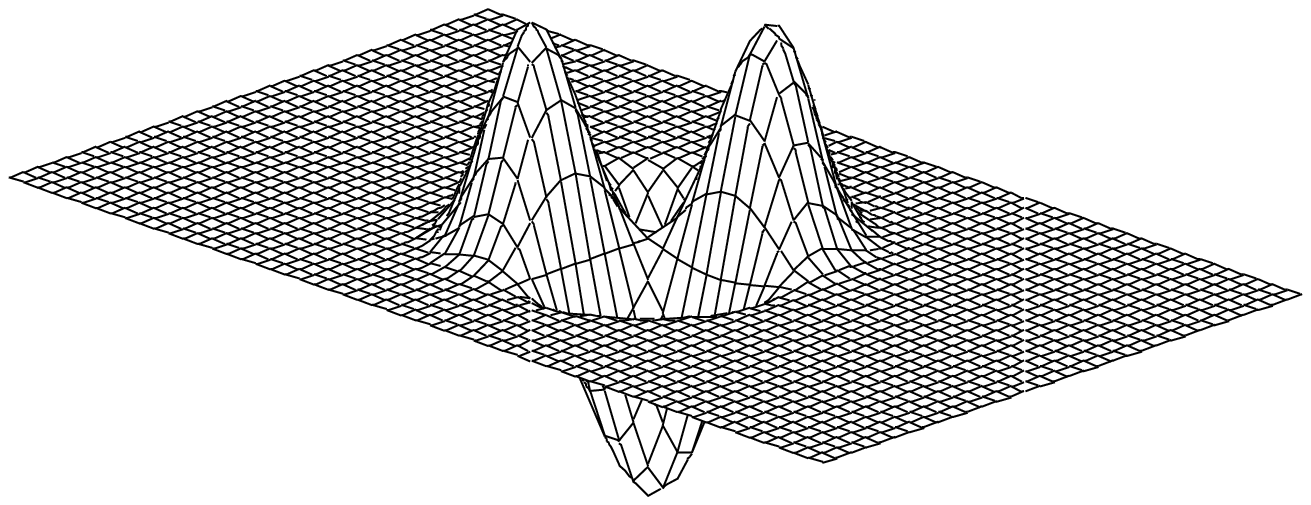}
\caption{Projector functions of the chlorine atom.
Top: two s-type projector functions, middle: p-type, bottom: d-type.}
\label{fig:2}
\end{figure}

In order to be able to apply the transformation operator to an
arbitrary auxiliary wavefunction, we need to be able to expand the
auxiliary wavefunction locally into the auxiliary partial waves.
\begin{eqnarray}
\tilde\Psi(\mathbf{r})=\sum_{i\in R} \tilde\phi_i(\mathbf{r})c_i
=\sum_{i\in R} \tilde\phi_i(\mathbf{r})
\langle\tilde{p}_i|\tilde\Psi\rangle  
\quad{\rm for}\quad |{\bf r}-{\bf R}_R|<r_{c,R}
\label{eq:ps1center}
\end{eqnarray}
which defines the projector functions $|\tilde{p}_i\rangle$.  The
projector functions probe the local character of the auxiliary wave
function in the atomic region.  Examples of projector functions are
shown in Figure~\ref{fig:2}.  From Eq.~\ref{eq:ps1center} we can derive
$\sum_{i\in R}|\tilde\phi_i\rangle\langle\tilde{p}_i|=1$, which is valid
within $r_{c,R}$.  It can be shown by insertion, that the identity
Eq.~\ref{eq:ps1center} holds for any auxiliary wavefunction
$|\tilde\Psi\rangle$ that can be expanded locally into auxiliary
partial waves $|\tilde\phi_i\rangle$, if
\begin{equation}
\langle\tilde{p}_i|\tilde\phi_j\rangle=\delta_{i,j} 
\quad\mathrm{for}\quad i,j\in R
\end{equation}
Note that neither the projector functions nor the partial waves need
to be orthogonal among themselves. The projector functions are fully
determined with the above conditions and a closure relation, which is
related to the unscreening of the pseudopotentials (see Eq.~90 in
\citep{PAW94}).

By combining Eq.~\ref{eq:sr1} and Eq.~\ref{eq:ps1center}, we can apply
${\cal S}_R$ to any auxiliary wavefunction.
\begin{eqnarray}
{\cal S}_R|\tilde\Psi\rangle
&=&\sum_{i\in R} {\cal S}_R |\tilde\phi_i\rangle\langle\tilde{p}_i|\tilde\Psi\rangle
=\sum_{i\in R} \Bigl(|\phi_i\rangle-|\tilde\phi_i\rangle\Bigr)
\langle\tilde{p}_i|\tilde\Psi\rangle
\end{eqnarray}

Hence the transformation operator is
\begin{eqnarray}
{\cal T}=1+\sum_i\Bigl(|\phi_i\rangle-|\tilde\phi_i\rangle\Bigr)
\langle\tilde{p}_i|
\label{eq:transf}
\end{eqnarray}
where the sum runs over all partial waves of all atoms.  The true wave
function can be expressed as
\begin{eqnarray}
|\Psi\rangle=|\tilde\Psi\rangle
+\sum_i\Bigl(|\phi_i\rangle-|\tilde\phi_i\rangle\Bigr)
\langle\tilde{p}_i|\tilde\Psi\rangle
=|\tilde\Psi\rangle
+\sum_R\Bigl( |\Psi^1_R\rangle-|\tilde\Psi^1_R\rangle\Bigr)
\label{eq:aewave}
\end{eqnarray}
with
\begin{eqnarray}
|\Psi^1_R\rangle&=&\sum_{i\in R}|\phi_i\rangle
\langle\tilde{p}_i|\tilde\Psi\rangle
\\
|\tilde\Psi^1_R\rangle&=&\sum_{i\in R}|\tilde\phi_i\rangle
\langle\tilde{p}_i|\tilde\Psi\rangle
\end{eqnarray}

\begin{figure}[h]
\centering
\includegraphics[height=10cm,angle=-90,clip=true]{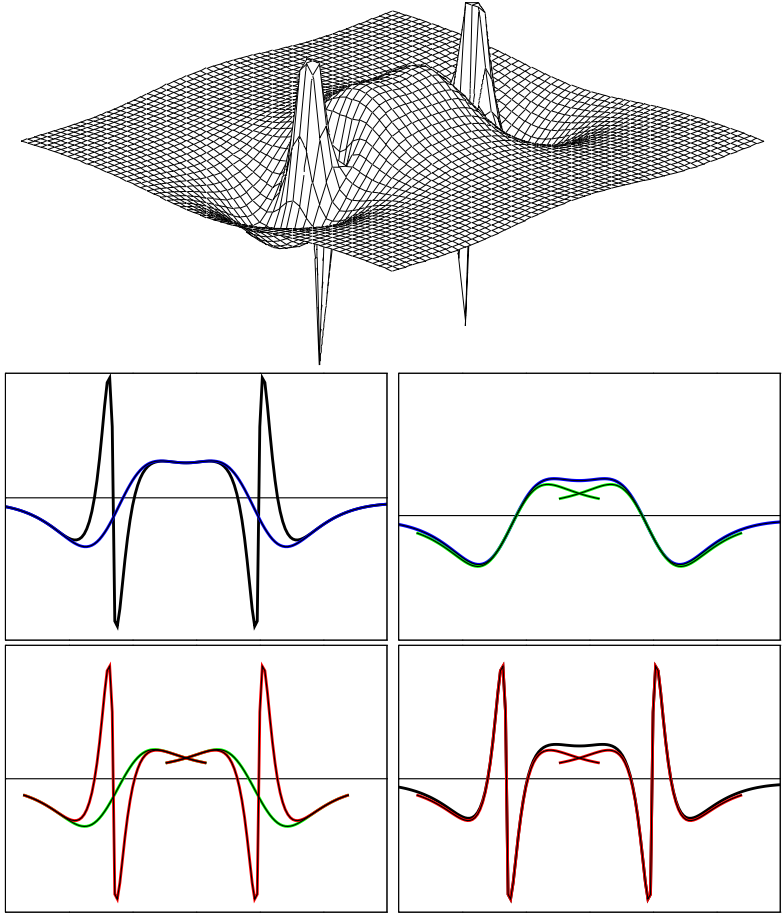}\\
\includegraphics[width=5.1cm,clip=true]{ae-ps_new}
\includegraphics[width=5.1cm,clip=true]{ps-ps1_new}\\
\includegraphics[width=5.1cm,clip=true]{ae1-ps1_new}
\includegraphics[width=5.1cm,clip=true]{ae-ae1_new}
\caption{Bonding p-$\sigma$ orbital of the Cl$_2$ molecule and its
decomposition of the wavefunction into auxiliary wavefunction and the
two one-center expansions.  Top-left: True and auxiliary wave
function; top-right: auxiliary wavefunction and its partial wave
expansion; bottom-left: the two partial wave expansions; bottom-right:
true wavefunction and its partial wave expansion.}
\label{fig:1}
\end{figure}

In Fig.~\ref{fig:1} the decomposition of Eq.~\ref{eq:aewave} is shown
for the example of the bonding p-$\sigma$ state of the Cl$_2$
molecule.

To understand the expression Eq.~\ref{eq:aewave} for the true wave
function, let us concentrate on different regions in space. (1) Far
from the atoms, the partial waves are, according to
Eq.~\ref{eq:equaloutside}, pairwise identical so that the auxiliary
wavefunction is identical to the true wavefunction, that is
$\Psi(\mathbf{r})=\tilde\Psi(\mathbf{r})$.  (2) Close to an atom $R$,
however, the auxiliary wavefunction is, according to
Eq.~\ref{eq:ps1center}, identical to its one-center expansion, that is
$\tilde\Psi(\mathbf{r})=\tilde\Psi^1_R(\mathbf{r})$.  Hence the true
wavefunction $\Psi(\mathbf{r})$ is identical to
$\Psi^1_R(\mathbf{r})$, which is built up from partial waves that
contain the proper nodal structure.

In practice, the partial wave expansions are truncated.  Therefore,
the identity of Eq.~\ref{eq:ps1center} does not hold strictly.  As a
result the plane-waves also contribute to the true wavefunction
inside the atomic region. This has the advantage that the missing
terms in a truncated partial wave expansion are partly accounted for
by plane-waves, which explains the rapid convergence of the partial
wave expansions. This idea is related to the additive augmentation
of the LAPW method of Soler \citep{Soler89}.

Frequently, the question comes up, whether the transformation
Eq.~\ref{eq:transf} of the auxiliary wavefunctions indeed provides
the true wavefunction. The transformation should be considered merely
as a change of representation analogous to a coordinate transform.  If
the total energy functional is transformed consistently, its minimum
will yield auxiliary wavefunctions that produce the correct wave
functions $|\Psi\rangle$.

%
\subsection{Expectation values}
Expectation values can be obtained either from the reconstructed
true wavefunctions or directly from the auxiliary wave
functions 
\begin{eqnarray}
\langle A\rangle&=&\sum_{n}f_n\langle\Psi_n|A|\Psi_n\rangle
+\sum_{n=1}^{N_c}\langle\phi_n^c|A|\phi_n^c\rangle
\nonumber\\
&=&\sum_{n}f_n\langle\tilde\Psi_n|{\cal T}^\dagger A{\cal T}
|\tilde\Psi_n\rangle
+\sum_{n=1}^{N_c}\langle\phi_n^c|A|\phi_n^c\rangle
\label{eq:expect1}
\end{eqnarray}
where $f_n$ are the occupations of the valence states and $N_c$ is the
number of core states. The first sum runs over the valence states, and
second over the core states $|\phi^c_n\rangle$.

Now we can decompose the matrix element for a wavefunction $\Psi$ into
its individual contributions according to Eq.~\ref{eq:aewave}.
\begin{eqnarray}
\langle\Psi|A|\Psi\rangle&=&
\langle\tilde\Psi+\sum_R(\Psi^1_R-\tilde\Psi^1_R)|
A|\tilde\Psi+\sum_{R'}(\Psi^1_{R'}-\tilde\Psi^1_{R'})\rangle
\nonumber\\
&=&\underbrace{\langle\tilde\Psi|A|\tilde\Psi\rangle
+\sum_R\Bigl(\langle\Psi^1_R|A|\Psi^1_R\rangle
-\langle\tilde\Psi^1_R|A|\tilde\Psi^1_R\rangle\Bigr)}_{\mbox{part 1}}
\nonumber\\
&+&\underbrace{\sum_R\Bigl(
\langle \Psi^1_R-\tilde\Psi^1_R|A|\tilde\Psi-\tilde\Psi^1_R\rangle
+\langle\tilde\Psi-\tilde\Psi^1_R|A|\Psi^1_R-\tilde\Psi^1_R\rangle
\Bigr)}_{\mbox{part 2}}
\nonumber\\
&+&\underbrace{\sum_{R\neq R'}\langle \Psi^1_R
-\tilde\Psi^1_R|A|\Psi^1_{R'}-\tilde\Psi^1_{R'}\rangle}
_{\mbox{part 3}}
\label{eq:expect}
\end{eqnarray}
Only the first part of Eq.~\ref{eq:expect}, is evaluated explicitly,
while the second and third parts of Eq.~\ref{eq:expect} are neglected,
because they vanish for sufficiently local operators as long as the
partial wave expansion is converged: The function
$\Psi^1_R-\tilde\Psi^1_R$ vanishes per construction beyond its
augmentation region, because the partial waves are pairwise identical
beyond that region. The function $\tilde\Psi-\tilde\Psi^1_R$ vanishes
inside its augmentation region, if the partial wave expansion is
sufficiently converged.  In no region of space both functions
$\Psi^1_R-\tilde\Psi^1_R$ and $\tilde\Psi -\tilde\Psi^1_R$ are
simultaneously nonzero.  Similarly the functions
$\Psi^1_R-\tilde\Psi^1_R$ from different sites are never non-zero in
the same region in space. Hence, the second and third parts of
Eq.~\ref{eq:expect} vanish for operators such as the kinetic energy
$\frac{-\hbar^2}{2m_e}\mathbf{\bnabla}^2$ and the real space projection
operator $|r\rangle\langle r|$, which produces the electron density.
For truly nonlocal operators the parts 2 and 3 of Eq.~\ref{eq:expect}
would have to be considered explicitly.

The expression, Eq.~\ref{eq:expect1}, for the expectation value can
therefore be written with the help of Eq.~\ref{eq:expect} as
\begin{eqnarray}
\langle A\rangle&=&
\sum_{n}f_n\Bigl(\langle\tilde\Psi_n|A|\tilde\Psi_n\rangle
+\langle\Psi^1_n|A|\Psi^1_n\rangle
-\langle\tilde\Psi^1_n|A|\tilde\Psi^1_n\rangle\Bigr)
+\sum_{n=1}^{N_c}\langle\phi^c_n|A|\phi^c_n\rangle
\nonumber\\
&=&\sum_{n}f_n\langle\tilde\Psi_n|A|\tilde\Psi_n\rangle
+\sum_{n=1}^{N_c}\langle\tilde\phi^c_n|A|\tilde\phi^c_n\rangle
\nonumber\\
&+&\sum_R\Bigl(\sum_{i,j\in R}D_{i,j}\langle\phi_j|A|\phi_i\rangle
+\sum_{n\in R}^{N_{c,R}}\langle\phi^c_n|A|\phi^c_n\rangle\Bigr)
\nonumber\\
&-&\sum_R\Bigl(\sum_{i,j\in R} D_{i,j}\langle\tilde\phi_j|A|\tilde\phi_i\rangle
+\sum_{n\in R}^{N_{c,R}}\langle\tilde\phi^c_n|A|\tilde\phi^c_n\rangle\Bigr)
\end{eqnarray}
where $D_{i,j}$ is the one-center density matrix defined as
\begin{eqnarray}
D_{i,j}=\sum_n
f_n\langle\tilde\Psi_n|\tilde{p}_j\rangle
\langle\tilde{p}_i|\tilde\Psi_n\rangle
=\sum_n \langle\tilde{p}_i|\tilde\Psi_n\rangle
f_n\langle\tilde\Psi_n|\tilde{p}_j\rangle
\label{eq:1cdenmat}
\end{eqnarray}

The auxiliary core states, $|\tilde\phi^c_n\rangle$ allow to
incorporate the tails of the core wavefunction into the plane-wave
part, and therefore assure, that the integrations of partial wave
contributions cancel strictly beyond $r_c$. They are identical to the
true core states in the tails, but are a smooth continuation inside
the atomic sphere. It is not required that the auxiliary wave
functions are normalized.

Following this scheme, the electron density is given by
\begin{eqnarray}
n(\mathbf{r})&=&\tilde{n}(\mathbf{r})
+\sum_R\Bigl(n^1_R(\mathbf{r})-\tilde{n}^1_R(\mathbf{r})\Bigr)
\\
\tilde{n}(\mathbf{r})
&=&\sum_n f_n \tilde\Psi^*_n(\mathbf{r})\tilde\Psi_n(\mathbf{r})+\tilde{n}_c
\nonumber\\
n^1_R(\mathbf{r})&=&\sum_{i,j\in R}
D_{i,j}\phi^*_j(\mathbf{r})\phi_i(\mathbf{r})+n_{c,R} \nonumber
\nonumber\\
\tilde{n}^1_R(\mathbf{r})&=&\sum_{i,j\in R}
D_{i,j}\tilde\phi^*_j(\mathbf{r})\tilde\phi_i(\mathbf{r})+\tilde{n}_{c,R}
\end{eqnarray}
where $n_{c,R}$ is the core density of the corresponding atom and
$\tilde{n}_{c,R}$ is the auxiliary core density that is identical to $n_{c,R}$
outside the atomic region, but smooth inside.

Before we continue, let us discuss a special point: The matrix element of a
general operator with the auxiliary wavefunctions may be slowly converging
with the plane-wave expansion, because the operator $A$ may not be well
behaved. An example for such an operator is the singular electrostatic
potential of a nucleus. This problem can be alleviated by adding an
``intelligent zero'': If an operator $B$ is purely localized within an atomic
region, we can use the identity between the auxiliary wavefunction and its
own partial wave expansion
\begin{eqnarray}
0&=&\langle\tilde\Psi_n|B|\tilde\Psi_n\rangle
-\langle\tilde\Psi_n^1|B|\tilde\Psi_n^1\rangle
\label{eq:zeroop}
\end{eqnarray}
Now we choose an operator $B$ so that it cancels the problematic
behavior of the operator $A$, but is localized in a single atomic
region.  By adding $B$ to the plane-wave part and the matrix elements
with its one-center expansions, the plane-wave convergence can be
improved without affecting the converged result. A term of this type,
namely $\bar{v}$ will be introduced in the next section to cancel the
Coulomb singularity of the potential at the nucleus.

\subsection{Total Energy}
Like wavefunctions and expectation values also the total
energy can be divided into three parts.
\begin{eqnarray}
E[\tilde\Psi_n,R_R]&=&\tilde{E}+\sum_R\Bigl(E^1_R-\tilde{E}^1_R\Bigr)
\end{eqnarray}
The plane-wave part $\tilde{E}$ involves only smooth functions and is
evaluated on equi-spaced grids in real and reciprocal space.  This part is
computationally most demanding, and is similar to the expressions in the
pseudopotential approach.
\begin{eqnarray}
\tilde{E}&=&\sum_n\langle\tilde\Psi_n|\frac{-\hbar^2}{2m_e}\mathbf{\bnabla}^2
|\tilde\Psi_n\rangle
\nonumber\\
&+&\frac{1}{2}\cdot\frac{e^2}{4\pi\epsilon_0}\int d^3r\int d^3r'
\frac{[\tilde{n}(\mathbf{r})+\tilde{Z}(\mathbf{r})]
[\tilde{n}(\mathbf{r}')+\tilde{Z}(\mathbf{r}')]}{|\mathbf{r}-\mathbf{r}'|}
\nonumber\\
&+&\int d^3r \bar{v}(\mathbf{r})\tilde{n}(\mathbf{r})
+E_{xc}[\tilde{n}(\mathbf{r})]
\label{eq:psetot}
\end{eqnarray}
$\tilde{Z}({\bf r})$ is an angular-momentum dependent core-like
density that will be described in detail below.  The remaining parts
can be evaluated on radial grids in a spherical harmonics expansion.
The nodal structure of the wavefunctions can be properly described on
a logarithmic radial grid that becomes very fine near nucleus,
\begin{eqnarray}
{E}^1_R&=&\sum_{i,j\in R} D_{i,j}
\langle\phi_j|\frac{-\hbar^2}{2m_e}\mathbf{\bnabla}^2|\phi_i\rangle
+\sum_{n\in R}^{N_{c,R}}
\langle\phi^c_n|\frac{-\hbar^2}{2m_e}\mathbf{\bnabla}^2|\phi^c_n\rangle
\nonumber\\
&+&\frac{1}{2}\cdot\frac{e^2}{4\pi\epsilon_0}\int d^3r\int d^3r'
\frac{[n^1(\mathbf{r})+Z(\mathbf{r})][n^1(\mathbf{r}')+Z(\mathbf{r}')]}
{|\mathbf{r}-\mathbf{r}'|}
\nonumber\\
&+&E_{xc}[ n^1(\mathbf{r})]
\\
\tilde{E}^1_R&=&\sum_{i,j\in R} D_{i,j}
\langle\tilde\phi_j|\frac{-\hbar^2}{2m_e}\mathbf{\bnabla}^2|\tilde\phi_i\rangle
\nonumber\\
&+&\frac{1}{2}\cdot\frac{e^2}{4\pi\epsilon_0}\int d^3r\int d^3r'
\frac{[\tilde{n}^1(\mathbf{r})+\tilde{Z}(\mathbf{r})]
[\tilde{n}^1(\mathbf{r}')+\tilde{Z}(\mathbf{r}')]}{|\mathbf{r}-\mathbf{r}'|}
\nonumber\\
&+&
\int d^3r \bar{v}(\mathbf{r})\tilde{n}^1(\mathbf{r})
+E_{xc}[\tilde{n}^1(\mathbf{r})]
\label{eq:ps1etot}
\end{eqnarray}

The compensation charge density
$\tilde{Z}(\mathbf{r})=\sum_R\tilde{Z}_R(\mathbf{r})$ is given as a
sum of angular momentum dependent Gauss functions, which have an
analytical plane-wave expansion.  A similar term occurs also in the
pseudopotential approach. In contrast to the norm-conserving
pseudopotential approach, however, the compensation charge of an atom
$\tilde{Z}_R$ is non-spherical and constantly adapts to the
instantaneous environment. It is constructed such that
\begin{eqnarray}
n^1_R(\mathbf{r})+Z_R(\mathbf{r}) -\tilde{n}^1_R(\mathbf{r})
-\tilde{Z}_R(\mathbf{r})
\end{eqnarray}

has vanishing electrostatic multi-pole moments for each atomic site.
With this choice, the electrostatic potentials of the augmentation
densities vanish outside their spheres. This is the reason that there is
no electrostatic interaction of the one-center parts between different
sites.

The compensation charge density as given here is still localized
within the atomic regions. A technique similar to an Ewald summation,
however, allows to replace it by a very extended charge density. Thus
we can achieve, that the plane-wave convergence of the total energy
is not affected by the auxiliary density.

The potential $\bar{v}=\sum_R\bar{v}_R$, which occurs in
Eqs.~\ref{eq:psetot} and~\ref{eq:ps1etot} enters the total energy in
the form of ``intelligent zeros'' described in Eq.~\ref{eq:zeroop}
\begin{eqnarray}
0=\sum_nf_n\left(\langle\tilde\Psi_n|\bar{v}_R|\tilde\Psi_n\rangle
-\langle\tilde\Psi^1_n|\bar{v}_R|\tilde\Psi^1_n\rangle\right)
=\sum_nf_n\langle\tilde\Psi_n|\bar{v}_R|\tilde\Psi_n\rangle
-\sum_{i,j\in R}D_{i,j}
\langle\tilde\phi_i|\bar{v}_R|\tilde\phi_j\rangle
\end{eqnarray}
The main reason for introducing this potential is to cancel the
Coulomb singularity of the potential in the plane-wave part. The
potential $\bar{v}$ allows to influence the plane-wave convergence
beneficially, without changing the converged result.  $\bar{v}$ must
be localized within the augmentation region, where
Eq.~\ref{eq:ps1center} holds.
%
\subsection{Approximations}
Once the total energy functional provided in the previous section has
been defined, everything else follows: Forces are partial derivatives
with respect to atomic positions.  The potential is the derivative of
the non-kinetic energy contributions to the total energy with respect
to the density, and the auxiliary Hamiltonian follows from derivatives
$\tilde{H}|\tilde\Psi_n\rangle$ with respect to auxiliary wave
functions.  The fictitious Lagrangian approach of Car and Parrinello
\citep{Car85} does not allow any freedom in the way these derivatives
are obtained. Anything else than analytic derivatives will violate
energy conservation in a dynamical simulation. Since the expressions
are straightforward, even though rather involved, we will not discuss
them here.

All approximations are incorporated already in the total energy functional of
the PAW method. What are those approximations?
\begin{itemize}
\item Firstly we use the frozen-core approximation. In principle this
approximation can be overcome.
\item The plane-wave expansion for the auxiliary wavefunctions must
be complete. The plane-wave expansion is controlled easily by
increasing the plane-wave cutoff defined as
$E_{PW}=\frac{1}{2}\hbar^2G_{max}^2$.  Typically we use a plane-wave
cutoff of 30~Ry.
\item The partial wave expansions must be converged. Typically we use
one or two partial waves per angular momentum $(\ell,m)$ and site.
It should be noted that the partial wave expansion is not
variational, because it changes the total energy functional and not
the basis set for the auxiliary wavefunctions.
\end{itemize}
We do not discuss here numerical approximations such as the choice of
the radial grid, since those are easily controlled.

\subsection{Relation to the Pseudopotentials}

We mentioned earlier that the pseudopotential approach can be derived
as a well defined approximation from the PAW method: The
augmentation part of the total energy $\Delta E=E^1-\tilde{E}^1$ for
one atom is a functional of the one-center density matrix $D_{i,j\in R}$
defined in Eq.~\ref{eq:1cdenmat}. The pseudopotential approach can be
recovered if we truncate a Taylor expansion of $\Delta E$ about the
atomic density matrix after the linear term.  The term linear to
$D_{i,j}$ is the energy related to the nonlocal pseudopotential.
\begin{eqnarray}
\Delta E(D_{i,j})&=&\Delta E(D_{i,j}^{at})
+\sum_{i,j}(D_{i,j}-D^{at}_{i,j})\frac{\partial \Delta E}{\partial D_{i,j}}
+O(D_{i,j}-D^{at}_{i,j})^2
\nonumber\\
&=&E_{self}+\sum_n f_n\langle\tilde\Psi_n|v^{ps}|\tilde\Psi_n\rangle
-\int d^3r \bar{v}(\mathbf{r})\tilde{n}(\mathbf{r})
+O(D_{i,j}-D^{at}_{i,j})^2
\end{eqnarray}
which can directly be compared to the total energy expression
Eq.~\ref{eq:totalenergypseudopotential} of the pseudopotential
method. The local potential $\bar{v}(\mathbf{r})$ of the
pseudopotential approach is identical to the corresponding potential
of the projector augmented wave method. The remaining contributions in
the PAW total energy, namely $\tilde{E}$, differ from the
corresponding terms in Eq.~\ref{eq:totalenergypseudopotential} only in
two features: our auxiliary density also contains an auxiliary core
density, reflecting the nonlinear core correction of the
pseudopotential approach, and the compensation density
$\tilde{Z}(\mathbf{r})$ is non-spherical and depends on the wave
function.  Thus we can look at the PAW method also as a
pseudopotential method with a pseudopotential that adapts to the
instantaneous electronic environment. In the PAW method, the explicit
nonlinear dependence of the total energy on the one-center density
matrix is properly taken into account.

What are the main advantages of the PAW method compared to the
pseudopotential approach?  

Firstly all errors can be systematically controlled so that there are
no transferability errors.  As shown by Watson \citep{Watson98} and
Kresse \citep{Kresse99}, most pseudopotentials fail for high spin
atoms such as Cr. While it is probably true that pseudopotentials can
be constructed that cope even with this situation, a failure can not
be known beforehand, so that some empiricism remains in practice: A
pseudopotential constructed from an isolated atom is not guaranteed to
be accurate for a molecule.  In contrast, the converged results of the
PAW method do not depend on a reference system such as an isolated
atom, because PAW uses the full density and potential.

Like other all-electron methods, the PAW method provides access to the
full charge and spin density, which is relevant, for example, for
hyperfine parameters. Hyperfine parameters are sensitive probes of the
electron density near the nucleus. In many situations they are the only
information available that allows to deduce atomic structure and
chemical environment of an atom from experiment.

The plane-wave convergence is more rapid than in norm-conserving
pseudopotentials and should in principle be equivalent to that of
ultra-soft pseudo\-potentials \citep{Vanderbilt90}. Compared to the
ultra-soft pseudo\-potentials, however, the PAW method has the
advantage that the total energy expression is less complex and can
therefore be expected to be more efficient.

The construction of pseudopotentials requires to determine a number of
parameters. As they influence the results, their choice is critical.
Also the PAW methods provides some flexibility in the choice of
auxiliary partial waves. However, this choice does not influence the
converged results.

\subsection{Recent Developments}

Since the first implementation of the PAW method in the CP-PAW
code, a number of groups have adopted the PAW method.  The
second implementation was done by the group of
Holzwarth  \citep{Holzwarth97}. The resulting PWPAW code is freely
available  \citep{Holzwarth01}.  This code is also used as a basis for
the PAW implementation in the AbInit project. An
independent PAW code has been developed by Valiev and
Weare  \citep{Valiev99}. Recently the PAW method has been implemented
into the VASP code  \citep{Kresse99}. The PAW method has also been
implemented by W. Kromen into the ESTCoMPP code of Bl\"ugel and
Schr\"oder.

Another branch of methods uses the reconstruction of the PAW method,
without taking into account the full wavefunctions in the energy
minimization. Following chemist notation this approach could be termed
``post-pseudopotential PAW''. This development began with the
evaluation for hyperfine parameters from a pseudopotential calculation
using the PAW reconstruction operator \citep{Hyperfine} and is now
used in the pseudopotential approach to calculate properties that
require the correct wavefunctions such as hyperfine parameters.

The implementation by Kresse and Joubert \citep{Kresse99} has been
particularly useful as they had an implementation of PAW in the same
code as the ultra-soft pseudo\-potentials, so that they could
critically compare the two approaches with each other. Their
conclusion is that both methods compare well in most cases, but they
found that magnetic energies are seriously -- by a factor two -- in
error in the pseudo\-potential approach, while the results of the PAW
method were in line with other all-electron calculations using the
linear augmented plane-wave method. As a short note, Kresse and
Joubert incorrectly claim that their implementation is superior as it
includes a term that is analogous to the non-linear core correction of
pseudo\-potentials \citep{Louie82}: this term however is already included
in the original version in the form of the pseudized core density.

Several extensions of the PAW have been done in the recent years:
For applications in chemistry truly isolated systems are often of
great interest. As any plane-wave based method introduces periodic
images, the electrostatic interaction between these images can cause
serious errors. The problem has been solved by mapping the charge
density onto a point charge model, so that the electrostatic
interaction could be subtracted out in a self-consistent
manner  \citep{decoupling}. In order to include the influence of the
environment, the latter was simulated by simpler force fields using
the molecular-mechanics-quantum-mechanics (QM-MM) approach  \citep{QMMM}.

In order to overcome the limitations of the density functional theory
several extensions have been performed. Bengone \citep{Bengone00}
implemented the LDA+U approach into the CP-PAW code.  Soon after this,
Arnaud \citep{Arnaud2000} accomplished the implementation of the GW
approximation into the CP-PAW code.  The VASP-version of PAW
\citep{Kresse00} and the CP-PAW code have now been extended to include
a non-collinear description of the magnetic moments. In a non-collinear
description the Schr\"odinger equation is replaced by the Pauli
equation with two-component spinor wavefunctions

The PAW method has proven useful to evaluate electric field
gradients  \citep{EFG} and magnetic hyperfine parameters with high
accuracy  \citep{SiO2}. Invaluable will be the prediction of NMR chemical
shifts using the GIPAW method of Pickard and Mauri  \citep{Mauri2001},
which is based on their earlier work  \citep{Mauri96}. While the GIPAW
is implemented in a post-pseudo\-potential manner, the extension to a
self-consistent PAW calculation should be straightforward.  An
post-pseudo\-potential approach has also been used to evaluate core
level spectra  \citep{Pickard97} and momentum matrix
elements  \citep{Kageshima}.

\begin{acknowledgments}
We are grateful for carefully reading the manuscript to S. Boeck,
J.\,Noffke, as well as to K.\,Schwarz for his continuous
support. This work has benefited from the collaborations within the
ESF Programme on 'Electronic Structure Calculations for Elucidating
the Complex Atomistic Behavior of Solids and Surfaces'.
\end{acknowledgments}


\end{document}